# Accurate potential energy, dipole moment curves, and lifetimes of vibrational states of heteronuclear alkali dimers

Dmitry A. Fedorov[a], Andrei Derevianko[b], Sergey A. Varganov[a,*]

[a]Department of Chemistry, University of Nevada, Reno, 1664 N. Virginia St., Reno, NV 89557-0216

[b]Department of Physics, University of Nevada, Reno, 1664 N. Virginia St. Reno, NV 89557-0220

*E-mail: svarganov@unr.edu

**Abstract**

We calculate the potential energy curves, the permanent dipole moment curves, and the lifetimes of the ground and excited vibrational states of the heteronuclear alkali dimers XY (X, Y = Li, Na, K, Rb, Cs) in the $X^1\Sigma^+$ electronic state using the coupled cluster with singles doubles and triples (CCSDT) method. All-electron quadruple-$\zeta$ basis sets with additional core functions are used for Li and Na, and small-core relativistic effective core potentials with quadruple-$\zeta$ quality basis sets are used for K, Rb and Cs. The inclusion of the coupled cluster non-perturbative triple excitations is shown to be crucial for obtaining the accurate potential energy curves. Large one-electron basis set with additional core functions is needed for the accurate prediction of permanent dipole moments. The dissociation energies are overestimated by only 14 cm$^{-1}$ for LiNa and by no more than 114 cm$^{-1}$ for the other molecules. The discrepancies between the experimental and calculated harmonic vibrational frequencies are less than 1.7 cm$^{-1}$, and the discrepancies for the anharmonic correction are less than 0.1 cm$^{-1}$. We show that correlation between atomic electronegativity differences and permanent dipole moment of heteronuclear alkali dimers is not perfect. To obtain the vibrational energies and wave functions the vibrational Schrödinger equation is solved with the B-spline basis set method. The transition dipole moments between all vibrational states, the Einstein coefficients and the lifetimes of the vibrational states are calculated. We analyze the decay rates of the vibrational states in terms of spontaneous emission, and stimulated emission and absorption induced by black body radiation. In all studied heteronuclear alkali dimers the ground vibrational states have much longer lifetimes than any excited states.

## I. INTRODUCTION

In recent years, ultracold molecules have become an active research field driven by several motivations, ranging from ultracold chemistry and attainment of degenerate quantum gases to quantum information processing. Other applications of ultracold molecules include precision measurements and interferometry. For example, the unprecedented control of chemical reactions can be achieved by changing the quantum states of reactants.[1-3] Trapped polar ultracold molecules can be used to simulate the quantum behavior of condensed matter systems, such as Hubbard models.[4] In an optical lattice polar molecules coupled via dipole-dipole interaction represent a system of interacting spins on a lattice. In such a system the transitions between different quantum phases can be simulated. Another possible application of ultracold molecules are measurements of fundamental physical constants, such as the electron dipole moment.[5,6] Ultracold diatomic molecules with nonzero dipole moments can be used as quantum bits. A possible scheme of a quantum computation device based on polar diatomic molecules has been proposed by DeMille.[7] In this scheme the Stark states of molecular dipole moments form qubit states. Molecules are trapped in a 1-dimensional optical lattice placed in a gradient of external electric field. The molecules are coupled via dipole-dipole interaction, forming a basis of multi-qubit operations, where single-qubit operations carried out using site-selective addressing with tunable microwave fields.

Ultracold alkali dimers are of a special interest as they can be produced by photoassociation of ultracold atoms, which have been efficiently cooled and trapped over the past two decades. The temperature of these molecules is expected to be comparable with the temperature of atoms, from which they are formed. Trapping and cooling of atoms is easier than that of molecules, because atoms do not have vibrational and rotational degrees of freedom. Several ultracold heteronuclear alkali dimers, such as RbCs,[8] KRb,[9,10] and NaCs[11,12] have been already obtained experimentally.

Accurate electronic structure calculations of the properties of heteronuclear alkali dimers can help in optimizing their production. Potential energy curves, vibrational energies and transition dipole moments can be useful in designing the future photoassociation experiments. In most of the applications listed above the ultracold molecules must stay in the same quantum state for a period of time required to perform measurements. The knowledge of the lifetimes of the vibrational quantum states can help in selecting the best molecule and vibrational state for specific application.

The early systematic theoretical study of heteronuclear alkali dimers was carried out by Aymar and Dulieu.[13] In this work the potential energy and permanent dipole moment curves for the ground singlet and the lowest triplet electronic states were computed using configuration interaction method in combination with core polarization potentials. The large-core polarization potentials were used to replace all the inner-shell electrons, effectively treating the alkali atoms as one-electron systems. Kotochigova *et al.*[14] evaluated properties of RbCs molecule with configuration interaction valence bond method. The permanent and transition dipole moments were calculated, and the effect of blackbody radiation on the molecule lifetime was estimated. Recently, the chemical reactions of ultracold alkali dimers in the lowest-energy $^3\Sigma^+$ state were studied using the restricted coupled cluster method with singles, doubles and perturbative triples (RCCSD(T)).[15]

The goal of this work is twofold: (1) to obtain the accurate spectroscopic constants, potential energy curves and permanent dipole moment curves of heteronuclear alkali dimers using modern high-level electronic structure methods; (2) to calculate the lifetimes of the ground and excited vibrational states of these dimers using the results of high-level electronic structure calculations.

The remainder of the paper is organized as follows. Section II describes the methodology used to solve the electronic and vibration Schrödinger equations, and to calculate the lifetimes of the ground and excited vibrational states. Section III consists of subsections A, B, C, D and E describing the results of electronic structure calculations on ten heteronuclear dimers XY (X, Y = Li, Na, K, Rb, Cs), and subsection F is dedicated to the lifetimes of the vibrational states of these molecules.

## II. COMPUTATIONAL METHODS

The electronic structure calculations were carried out using the following versions of the single reference coupled cluster method of increasing complexity: coupled cluster with singles and doubles (CCSD); with singles, double and perturbative triples (CCSD(T)); and with non-perturbative triples (CCSDT). The reference wave functions were obtained from the restricted Hartree-Fock method. To determine if a single reference coupled cluster method can compensate for the deficiency of the restricted Hartree-Fock wave function at large internuclear separation, we also performed the multireference configuration interaction with singles and doubles (MRCISD) calculations. The MRCISD reference wave function was of the CASSCF type with the full valence active space of two electrons on two ($\sigma$, $\sigma^*$) molecular orbitals. The MRCISD, CCSD and CCSD(T) calculations were done using quantum chemistry package MOLPRO.[16] The CCSDT calculations were performed with the CFOUR package.[17]

The cc-pVXZ, aug-cc-pVXZ and cc-pCVXZ (X = D, T, Q, 5) correlation consistent basis sets of Dunning *et al.*[18] were used for Li and Na atoms. For K, Rb and Cs the small-core relativistic effective core potentials (ECP) of Stuttgart/Cologne group were used.[19] These ECP leave one valence electron in the outer ns orbital, and 8 electrons in the (n-1) sp-shell, where n = 4, 5, 6 for K, Rb and Cs, respectively. In the coupled cluster and MRCISD calculations all electrons of Li and Na, and all 9 explicit electrons of K, Rb and Cs were correlated.

The equilibrium internuclear distance ($R_e$) and the permanent dipole moment curves ($\mu(R)$) were calculated with the CCSDT method. The dissociation energy was computed as the difference between the energy at the internuclear distance $R=R_e$ and the energy at the largest calculated internuclear distance ($R=28.3a_0$). The sign of the dipole moment depends on the orientation of interatomic axis. In this paper we use the following convention. The interatomic axis is directed from the lighter nucleus to the heavier one. The negative values of dipole moment indicate the excess of electron density on the lighter atom. However, at equilibrium internuclear distances we report the absolute values of permanent dipole moment $\mu_e$.

The molecules are assumed to be in the $X^1\Sigma^+$ electronic state and the $J=0$ rotational state. To obtain the vibrational energies and wave functions the nuclear vibrational Schrödinger equation (1) was solved using the B-spline basis set method.[20]

$$-\frac{1}{2m}\psi_v''(R) + V(R)\psi_v(R) = E_v\psi_v(R) \qquad (1)$$

Here $m$ is the reduced mass of a molecule and $V(R)$ is the Born-Oppenheimer potential obtained from the CCSDT calculations. The vibrational wave functions $\psi_v$ were expanded in B-spline basis set, and variational method was used to calculate the expansion coefficients. Using the permanent dipole moment function $\mu(R)$, obtained from the CCSDT calculations, and the vibrational wave functions we calculated the transition dipole moments $\langle i|\mu(R)|f\rangle$ between initial (*i*) and final (*f*) vibrational states.

The lifetime $\tau_i$ of vibrational state *i* was calculated as

$$\tau_i^{-1} = \sum_f A_{if} + \sum_f B_{if}, \qquad (2)$$

with the first summation running over vibrational states lying below the state *i*, and the second summation running over all states. The Einstein coefficient $A_{if}$, describing the probability of spontaneous emission from vibrational state *i* to the lower-energy state *f* reads

$$A_{if} = \frac{4\omega_{if}^3}{3c^3}\left|\langle i|\mu(R)|f\rangle\right|^2. \qquad (3)$$

Here $\omega_{if} = |E_f - E_i|$ is the transition frequency between states *i* and *f*. The black body radiation (BBR) coming from the surrounding environment at *T*=300 K can induce stimulated absorption and emission processes, which are described by the Einstein coefficient $B_{if}= A_{if}N(\omega_{if})$, where the number of black body photons is

$$N(\omega_{if}) = \left(\exp\left(\frac{\omega_{if}}{k_B T}\right) - 1\right)^{-1}. \qquad (4)$$

The harmonic vibrational frequency $\omega_e$ and the anharmonic correction $\omega_e\chi_e$, related to the transition frequency from the ground to the first excited state as $\omega_{01} = \omega_e - 2\omega_e\chi_e$, were calculated from the system of two equations obtained by expanding the two lowest vibration energies in Taylor series

$$E_v = \omega_e\left(v+\frac{1}{2}\right) - \omega_e\chi_e\left(v+\frac{1}{2}\right)^2, \tag{5}$$

where vibrational quantum number v = 0, 1.

### III. RESULTS AND DISCUSSION

The calculated and experimental values of the spectroscopic constants for ten heteronuclear alkali dimers are summarized in Table I. The harmonic vibrational frequencies and anharmonic corrections are presented in Table II.

#### A. LiNa molecule

We used the LiNa molecule as a benchmark to estimate the accuracy of electronic structure methods and the completeness of one-electron basis sets. First, the permanent dipole moment at the equilibrium distance was calculated using the CCSD method with the cc-pVXZ, aug-cc-pVXZ and cc-pCVXZ (X = D, T, Q, 5) basis sets (Fig. 1). The addition of diffuse functions (aug-cc-pVXZ basis sets) has no significant effect on the calculated values. In contrast, the core functions (cc-pCVXZ basis sets) must be included in order to obtain accurate permanent dipole moment. For example, at the equilibrium distance the difference between the experimental and cc-pVQZ values of the dipole moment is almost three times larger than the difference between the experimental and cc-pCVQZ values. It is important to notice that the dipole moment values computed with the cc-pCVXZ basis sets systematically converge to the experimental value with increasing size of the basis set. In contrast, without additional core functions (cc-pVXZ basis sets) the dipole moment convergence seems to be erratic. Halkier et al[21] showed that the dipole moment converges as $\mu_X = \mu_{\lim} + aX^{-3} + bX^{-4}$, where $a$ and $b$ are fitting coefficients, and $\mu_X$ and $\mu_{\lim}$ are the dipole moment values obtained with $X$-zeta basis set and in the complete basis set limit, respectively. We used the values obtained with the cc-pCVXZ basis set family to extrapolate the dipole moment and obtained $\mu_{\lim}$=0.454 D (Fig. 1), which is in an excellent agreement with the experimental value of $\mu_e$=0.45 D reported by Tarnovsky et al,[22] and with the same value from the laser induced fluorescence spectroscopy experiment by Engelke et al.[23] We did not do the extrapolation with CCSDT method because this would require the computationally expensive CCSDT/cc-pV5Z calculation, however, we expect the CCSDT basis set convergence to be similar to the CCSD one. The number of the basis functions in the basis set used with the Stuttgart/Cologne ECP is similar to the number of non-core basis functions in the cc-pCVQZ basis set. Therefore, to avoid basis set superposition error we used the cc-pCVQZ basis set for Li and Na in all following calculations. As can be seen from Fig. 1, even for such a large basis set the error in the dipole moment can be as large as 0.08 D. Therefore, the basis set incompleteness is expected to be the main source of error in our dipole moment calculations.

In the recent Fourier-transform spectroscopy study of the LiNa ground state by Steinke et al.[24] the dissociation energy $D_e$=7103.5±0.1 cm$^{-1}$ was obtained. This should be compared with our CCSDT/cc-pCVQZ value of 7117 cm$^{-1}$, which overestimates the experimental value by less than 14 cm$^{-1}$ (Table I). The experimental equilibrium distance $R_e$=5.4586$a_0$ is reasonably close to our value of 5.472$a_0$. The computed harmonic vibrational frequency of LiNa molecule, 257.4 cm$^{-1}$, is only 0.4 cm$^{-1}$ larger than the experimental value of 256.99 cm$^{-1}$.[23] In Fig. 2 we compare the potential energy curves of LiNa calculated with different electronic structure methods using the cc-pCVQZ basis set. MRCISD underestimates the dissociation energy by more than 2000 cm$^{-1}$, whereas CCSD overestimates it by almost 1000 cm$^{-1}$. Adding perturbative triples, as in the CCSD(T) method, results in more accurate dissociation energy, but produces an unphysical transition state at the internuclear distance of 2.1$R_e$.

It is worth noticing that, although the CCSD method with a large basis set is capable of producing accurate dipole moment values at the internuclear distances close to $R_e$, converging the CCSD amplitude equations at the internuclear distances longer than 3$R_e$ can be very challenging, if possible at all. The values of the permanent dipole moment at large internuclear distances are needed to compute the transition dipole moments matrix elements involving high vibrational levels. Therefore, for accurate prediction of the lifetimes of the vibrational excited states the potential energy and permanent dipole moment curves have to be calculated with at least the CCSDT method and quadruple-$\zeta$ quality basis sets. We used this level of theory in the calculations on all other heteronuclear alkali dimers studied in this work.

**B. LiX (X = K, Rb, Cs) molecules**

The LiK molecule was studied experimentally by Engelke *et al.*[25] Our value of equilibrium distance, $R_e$=6.273$a_0$, is very close to their experimental $R_e$=6.28$a_0$. The experimental dipole moment value, $\mu_e$=3.45±0.1 D, is also in good agreement with our computed value of 3.41 D. The dissociation energy, $D_e$=6150±120 cm$^{-1}$, reported in the same study, has large uncertainty and is somewhat below the calculated value of 6277 cm$^{-1}$. Aymar and Dulieu[13] in their calculations with large-core polarization potential obtained shorter equilibrium distance of 6.21$a_0$ and significantly larger dipole moment of 3.56 D.

In the recent Fourier-transform spectroscopy study of LiRb molecule by Ivanova *et al.*[26] the equilibrium distance of $R_e$=6.550$a_0$ was obtained, which is in a good agreement with our calculated value of 6.554$a_0$. The experimental value of the dissociation energy, $D_e$=5927.9±0.4 cm$^{-1}$, is overestimated by 74 cm$^{-1}$ in our calculations. Our value of the dipole moment, $\mu_e$=4.06 D, agrees with experimental value of 4.0±0.1 D reported by Tarnovsky.[22]

The LiCs molecule was studied using Fourier transform spectroscopy by Staanum *et al.*[27] The reported equilibrium distance is $R_e$=6.9317$a_0$, which is slightly shorter than our value of 6.945$a_0$. In the same experiment the dissociation energy of 5875.455±0.1 cm$^{-1}$ was obtained, which is overestimated in our calculations by 78 cm$^{-1}$. The dipole moment, $\mu_e$=5.5±0.2 D, measured by Deiglmayr *et al.*[28] agrees well with our calculated value of 5.36 D. The LiCs molecule was also studied theoretically by Sørensen *et al.*[29] using the spin-free relativistic CCSDT method with all-electron uncontracted ANO-RCC basis set. They obtained the dipole moment value of 5.447 D at the equilibrium distance of 6.9632$a_0$. Their dissociation energy, $D_e$=5835 cm$^{-1}$, is 40 cm$^{-1}$ below the experimental value. The discrepancies in their and our CCSDT values must be attributed to the different basis sets and the different treatment of relativistic effects.

The potential energy curves for the LiX (X = K, Rb, Cs) series of alkali dimers are presented in Fig. 3a. As the reduced mass of the molecules in the series increases, the depth of the potential well decreases and the equilibrium distance increases. For the molecules in the LiX series the dissociation energy is overestimated by 14-78 cm$^{-1}$ in comparison with experimental data.[24-27]

Fig. 3b shows the dipole moment curves for the LiX series with the experimental and calculated values of the dipole moment at the equilibrium distance indicated by squares and diamonds, respectively. For all LiX molecules, with exception of LiNa, the theoretical and experimental values of dipole moment agree within experimental error.

We analyzed the correlation between the magnitude of the dipole moment and the difference in the electronegativities of the atoms in the molecules. The difference in electronegativity between atoms A and B is calculated according to the Pauling's definition as $|X_A - X_B| = \sqrt{D_{AB} - (D_{A_2} D_{B_2})^{1/2}}$, where $D_{AB}$, $D_{A_2}$, $D_{B_2}$ are the dissociation energies of the diatomic molecules AB, A$_2$, and B$_2$, respectively. For the LiX series the electronegativity increases from 0.05 (LiNa) to 0.19 (LiCs), and the dipole moment monotonically increases from 0.54 D (LiNa) to 5.36 D (LiCs) indicating strong correlation between the Pauling's electronegativity difference and the magnitude of the dipole moment.

The calculated harmonic vibrational frequencies and anharmonic corrections together with experimental and earlier theoretical values[13] are reported in Tables I and II. For all the molecules in the LiX series the calculated harmonic vibrational frequencies are within 1.2 cm$^{-1}$ of the experimental values. The calculated anharmonic correction for LiK molecule is only 0.1 cm$^{-1}$ lower than the experimental value of 1.224 cm$^{-1}$.[25] For comparison, the previously reported theoretical values of the harmonic vibrational frequencies for LiRb and LiCs molecules[13] are 10 cm$^{-1}$ and 19 cm$^{-1}$ below the experimental values.

**C. NaX (X = K, Rb, Cs) molecules**

In the Fourier-transform spectroscopy experiment by Gerdes *et al.*[30] the $R_e$=6.6121$a_0$ and $D_e$=5273.62±0.10 cm$^{-1}$ were obtained for the NaK molecule. Our value of the dissociation energy, $D_e$=5364 cm$^{-1}$, at the equilibrium distance $R_e$=6.622$a_0$, is 90 cm$^{-1}$ above the experimental value. The calculated dipole moment value, $\mu_e$=2.68 D, agrees with experimental value of 2.76±0.1 D reported by Dagdigian *et al.*[31] within the experimental error.

According to the spectroscopic study of Wang *et al.*[32] the NaRb molecule has the equilibrium distance $R_e$=6.8849$a_0$, which is close to our value of 6.903$a_0$. Our dissociation energy, $D_e$=5128 cm$^{-1}$, is larger than the experimental value, $D_e$=5030.75±0.1 cm$^{-1}$, by 97 cm$^{-1}$. The dipole moment measured by Dagdigian *et al.*[31], 3.1±0.3 D, is in agreement with our value of 3.29 D. The calculated harmonic frequency $\omega_e$=106.0 cm$^{-1}$ is within 1 cm$^{-1}$ of the experimental value of 106.965 cm$^{-1}$. The calculated anharmonic correction, 0.3 cm$^{-1}$, is in a good agreement with the experimental value of 0.3636 cm$^{-1}$.[33]

The reported equilibrium distance for the NaCs molecule, $R_e$=7.2754$a_0$, obtained from Doppler spectroscopy by Diemer *et al.*[34], is somewhat lower than our value of 7.301$a_0$. In the work of Dagdigian *et al.*[31] the dipole moment of NaCs is reported to be 4.75±0.20 D, which is slightly above our value of 4.53 D. The experimental dissociation energy, $D_e$=4954.237±0.1 cm$^{-1}$, obtained by Docenko *et al.*[35], is overestimated by 111 cm$^{-1}$ in our calculations.

The same trends that were observed in the properties of the LiX molecules can be seen in the NaX molecules (Fig. 4). The equilibrium distances are larger and the dissociation energies are lower for the heavier molecules. The harmonic vibrational frequencies of NaK, NaRb and NaCs are lower than experimental values by 1.7, 1.0 and 1.2 cm$^{-1}$, respectively. The dipole moment increases from 0.54 D (LiNa) to 5.36 D (NaCs) as the electronegativity difference between atoms in a molecule increases from 0.05 (LiNa) to 0.14 (NaCs).

**D. KRb and KCs molecules**

The ultracold gas of KRb molecules was obtained by Ni *et al.*[10] The dipole moment, $\mu_e$=0.566±0.017 D, was measured using the dc Stark spectroscopy. The experimental equilibrium distance is $R_e$=7.69$a_0$.[36] The calculations of the dipole moment of KRb was performed by Kotochigova *et al.*[37] who obtained $\mu_e$=0.76 D and $R_e$=7.7$a_0$, and by M. Aymar and O. Dulieu[13] who predicted $\mu_e$=0.62 D and $R_e$=7.64$a_0$. In our CCSDT calculations we obtained the dipole moment of 0.65 D at the equilibrium distance of 7.699$a_0$, which is 0.08 D larger than the experimental value. The dissociation energy of KRb, $D_e$=4217.91±0.42 cm$^{-1}$, was estimated[38] based on previous experimental data of Kasahara *et al.*[39] and Wang *et al.*[40] Our calculated dissociation energy is 88 cm$^{-1}$ larger than this estimated value. The calculated harmonic vibrational frequency of 75.3 cm$^{-1}$ is in excellent agreement with the experimental value of 75.5 cm$^{-1}$ reported by Cavaliere *et al.*[41]

As to KCs, in the Fourier-transform spectroscopy study by Ferber *et al.*[42] the equilibrium distance, $R_e$=8.0955$a_0$, and the dissociation energy, $D_e$=4069.3±1.5 cm$^{-1}$, were obtained, which are close to our values of 8.111$a_0$ and 4183 cm$^{-1}$. However, the calculated value of the dipole moment, $\mu_e$=1.90 D, differs significantly from the experimental value of 2.58±0.3 D reported by Tarnovsky.[22] Because our value is in agreement with the theoretical prediction by Aymar and Dulieu[13] the experimental value can be inaccurate. The calculated harmonic frequency of 67.8 cm$^{-1}$ is lower than the experimental value[42] by 0.6 cm$^{-1}$. The predicted anharmonic correction, 0.2 cm$^{-1}$, is in an excellent agreement with the experimental value of 0.193 cm$^{-1}$.[42]

The potential energy curves for the KX series are shown in Fig. 5a. The same trend of decreasing dissociation energy with increasing reduced mass of a molecule is present. The dipole moment curves $\mu(R)$ are shown in Fig. 5b. The dipole moment values at the equilibrium distance agree well with the experimental values for all molecules except KCs, for which the experimental value can be inaccurate. The largest discrepancy between our calculated and experimental harmonic vibrational frequencies of the KX molecules is 1.7 cm$^{-1}$ (NaK).

**E. RbCs molecule**

The experimentally determined equilibrium internuclear distance of RbCs, $R_e$=8.26$a_0$, is significantly smaller than our calculated value of 8.380$a_0$. In the spectroscopic study by Fellows *et al*[43] the dissociation energy, $D_e$=3836.1±0.5 cm$^{-1}$, was reported. Our calculations overestimate the experimental value by 68 cm$^{-1}$. The experimental value of the dipole moment, $\mu_e$=1.3±0.1 D, reported by Kato et al[44] agrees with our calculated value of 1.21 D within the experimental error. The dipole moment calculated by Kotochigova et al.,[14] $\mu_e$=1.25 D, is also in a good agreement with our result. Our harmonic vibration frequency, 49.7 cm$^{-1}$, is in an excellent agreement with the experimental value.

The potential energy and dipole moment curves for RbX series are presented in Fig. 6. The potential energy curves follow the same trend as in the other series of decreasing $D_e$ and increasing $R_e$ with increasing reduced mass of a molecule. From the dipole moment curves it can be seen that the larger the electronegativity difference between the atoms in a molecule the larger the value of dipole moment.

**F. Vibrational state lifetimes**

The vibrational state lifetimes of alkali dimers were calculated using the transition frequencies and dipole moments obtained from the CCSDT potential energy and permanent dipole moment curves (eq. 1-4). The lifetimes of the ground vibrational states depends on the BBR absorption rate only, and therefore require the Einstein coefficients $B_{0f}$, where the subscript $f$ runs over all excited states. Our calculated ground state lifetimes, and the lifetimes obtained by Vanhaecke and O. Dulieu using the harmonic approximation for potential energy curves,[45] are presented in Table III. The ground vibrational state lifetimes vary from 53 s for LiCs to 7.3×10$^4$ s for RbCs. For all molecules, with exception of RbCs, the harmonic approximation values are larger because the lifetime is affected by ν=0→1 transition only, and the transitions to the higher excited states are strictly forbidden. In contrast, in our calculations done using actual anharmonic potential the transitions to higher excited vibrational states also contribute to the lifetime reduction. The slightly smaller lifetime of RbCs calculated using harmonic approximation can be explained by the differences in the potential energy and dipole moments curves used in the two studies.

As can be seen from eq. (3) the lifetime of the ground vibrational state strongly depends on the vibrational frequency and transition dipole moment. The results in Table III and in Fig. 3-6 indicate that the lifetime of the ground vibrational states is mostly determined by the magnitude of the permanent dipole moment, which contributes to the transition dipole moments. In general, the molecules with larger permanent dipole moments have shorter lifetime than the ones with smaller dipole moments. For example, the lifetime of the ground vibrational states of the LiX (X = Na, K, Rb, Cs) molecules monotonically decrease from $1.3 \times 10^3$ s (LiNa) to 53 s (LiCs) as the permanent dipole moments $\mu_e$ decrease from 0.54 D (LiNa) to 5.36 D (LiCs). However, the influence of the vibration frequency on the lifetime is also evident, for example, from the comparison of LiNa and KRb molecules. The permanent dipole moment of LiNa (0.54 D) is smaller than the dipole moment of KRb (0.65 D), but the lifetime of the ground vibrational states of LiNa is shorter ($1.3 \times 10^3$ s) than the lifetime of KRb ($5.6 \times 10^4$ s). The shorter lifetime of LiNa is due to its much larger vibrational frequency (257.4 cm$^{-1}$) compared to the KRb vibrational frequency (75.3 cm$^{-1}$).

The lifetimes of all vibrational states of the LiX (X = Na, K, Rb, Cs) molecules as functions of the vibrational quantum number ν are shown in Fig 7. The general trends in the lifetimes of the vibrational excited states of heteronuclear alkali dimers can be explained using the LiNa molecule as an example. Initially, the lifetime of the vibrational states decreases very rapidly with ν. This is because the excited states can spontaneously decay to the ground state or to the lower excited states, whereas the ground state can only absorb black body photons. From Fig. 8 it can be seen that the contribution from spontaneous decay to the transition rate is much larger than the contribution from stimulated absorption and emission. In the next region (ν=4-28) the decay becomes slower due to the smaller transition dipoles moments between neighboring states, which is related to the increase in anharmonicity of the potential energy curve. The following region (ν=29-39), where lifetime decays faster again, corresponds to large increase in spontaneous transition rate (Fig. 8). This increase comes from the large transition dipole moments between the state of interest (*i*) and lower-energy states (*f*), as shown on the insert of Fig. 8. Finally, after reaching the minimum at ν=39 the lifetime start increasing again. This final lifetime increase takes place because the transition dipole moments between the last few vibrational states and the lower-energy states are relatively small (see Fig. 8 insert). The vibrational state lifetimes as functions of the vibrational quantum number for other heteronuclear alkali dimers behave similarly to the lifetimes of the LiX vibrational states. The corresponding figures for other molecules studied in this work are available in the supplemental material.

**IV. SUMMARY**

We used the coupled cluster method with singles, doubles and non-perturbative triples (CCSDT), and quadruple-$\zeta$ quality basis sets to calculate the potential energy and dipole moment curves for heteronuclear alkali dimers. These potential energy and dipole moment curves allowed us to evaluate lifetimes of the ground and excited vibrational states.

Within the LiX, NaX, KX and RbX series the dissociation energy decreases and the equilibrium internuclear distance increases with increasing reduced mass of the molecule (Fig. 3a-6a). The potential energy curve with the largest dissociation energy ($D_e$=7117 cm$^{-1}$) and vibrational frequency ($\omega_e$=257.4 cm$^{-1}$), and the shortest equilibrium distance ($R_e$=5.472$a_0$) corresponds to the LiNa molecule. The potential energy curve with the smallest dissociation energy ($D_e$=3904 cm$^{-1}$) and vibrational frequency ($\omega_e$=49.7 cm$^{-1}$), and the longest equilibrium distance ($R_e$=8.380$a_0$) corresponds to the RbCs molecule.

We emphasize that obtaining accurate potential energy curves for the ground electronic state of the alkali dimers requires including non-perturbative triples in the coupled cluster calculations. The CCSDT dissociation energies are overestimated by only 14 cm$^{-1}$ for LiNa (best case) and by 114 cm$^{-1}$ for KCs (worst case). In contrast, MRCISD underestimates the dissociation energy of LiNa by more than 2000 cm$^{-1}$, CCSD overestimates it by almost 1000 cm$^{-1}$, and CCSD(T) produces qualitatively incorrect potential energy curve with non-physical transition state. The large difference in the dissociation energy errors of the LiNa and heavier molecules is probably due to the use of relativistic ECP. The LiNa is the only molecule, which was studied with all-electron basis set, without using ECP. Therefore, in principle, it should be possible to improve the accuracy of the potential energy curves of the heavier molecules by doing all-electron calculations with relativistic Breit-Pauli or Douglas-Kroll Hamiltonians, and a large basis set.

The general trends in the magnitude of permanent dipole moment can be explained using the Pauling's electronegativity difference. Within each LiX, NaX, KX and RbX series the permanent dipole moment increases with increasing difference in the electronegativity of the atoms in a molecule (Table III). The molecules assembled from atoms with the smallest electronegativity difference, such as LiNa, KRb, have the lowest values of the dipole moment. The largest permanent dipole moment (5.36 D) corresponds to the LiCs molecule with the highest value of the electronegativity difference (0.19). However, when molecules from different series are compared, this correlation does not always hold. For example, the LiK and LiRb molecules have lower dipole moments, but higher values of the electronegativity difference than NaCs.

We employed the CCSDT potential energy curves in calculations of vibrational wave functions and vibrational energies for ground and excited states by solving the vibrational Schrödinger equation. The transition dipole moments between all vibrational states were calculated using the CCSDT permanent dipole moment curves and vibrational wave functions. The calculated ground state BBR-induced lifetimes vary significantly from 53 s for LiCs to $7.3 \times 10^4$ s for RbCs. In general, molecules with large permanent dipole moments and vibrational frequencies have smaller ground vibrational state lifetimes. For all studied alkali dimers the ground vibrational state has the largest lifetime. Lifetime of the excited states as a function of the vibrational quantum number initially rapidly decreases, mostly due to spontaneous emission. However, after reaching a minimum, the lifetime slightly increases before reaching the last excited vibrational state. We expect that the calculated vibrational energies and vibrational state lifetimes will be useful in the future experiments involving ultracold heteronuclear alkali dimers.

See supplemental material at [URL will be inserted by AIP Publishing] for the CCSDT values of the potential energy and permanent dipole moment at different internuclear distances, the energies and lifetimes of the vibrational states, and the values of the transition dipole moments $\langle 0|\mu(R)|f\rangle$.

**ACKNOWLEDGMENTS**

We are grateful to University of Nevada, Reno for financial support in the form of the start-up fund for S.A.V., and the Gene and Clara LeMay Scholarship for D.A.F. Work of A.D. was supported in part by the NSF grant PHY-1306343. We thank Professor Olivier Dulieu for illuminating discussions.

**Table I.** Equilibrium distances ($R_e$), permanent dipole moments ($\mu_e$), dissociation energies ($D_e$), and harmonic vibrational frequencies ($\omega_e$) of the XY (X, Y = Li, Na, K, Rb, Cs) heteronuclear alkali dimers. The superscripts next to the experimental values indicate the references from which these values were taken.

| | This Work | | | | Experiment | | | | Theory (Ref.13) | |
|---|---|---|---|---|---|---|---|---|---|---|
| | $R_e$, $a_0$ | $\mu_e$, D | $D_e$, cm$^{-1}$ | $\omega_e$, cm$^{-1}$ | $R_e$, $a_0$ | $\mu_e$, D | $D_e$, cm$^{-1}$ | $\omega_e$, cm$^{-1}$ | $R_e$, $a_0$ | $\mu_e$, D |
| LiNa | 5.472 | 0.54 | 7117 | 257.4 | 5.4586[24] | 0.45(1)[22] | 7103.5(1)[24] | 256.99[23] | 5.43 | 0.56 |
| LiK | 6.273 | 3.41 | 6277 | 212.4 | 6.28[25] | 3.45(10)[31] | 6150(120)[25] | 211.92(2)[25] | 6.21 | 3.56 |
| LiRb | 6.554 | 4.06 | 6002 | 196.2 | 6.550[26] | 4.0(1)[22] | 5927.9(4)[26] | 195[47] | 6.52 | 4.17 |
| LiCs | 6.945 | 5.36 | 5953 | 182.6 | 6.9317[27] | 5.5(2)[48] | 5875.455(100)[27] | 183[47] | 6.81 | 5.52 |
| NaK | 6.622 | 2.68 | 5364 | 122.4 | 6.6121[30] | 2.76(10)[31] | 5273.62(10)[30] | 124.13[49] | 6.50 | 2.76 |
| NaRb | 6.903 | 3.29 | 5128 | 106.0 | 6.8849[32] | 3.1(3)[31] | 5030.75(10)[50] | 106.965[33] | 6.84 | 3.30 |
| NaCs | 7.301 | 4.53 | 5065 | 97.8 | 7.27[34] | 4.75(20)[31] | 4954.237(100)[35] | 99[34] | 7.20 | 4.61 |
| KRb | 7.699 | 0.65 | 4306 | 75.3 | 7.69[36] | 0.566(17)[10] | 4217.91(42)[38] | 75.5[41] | 7.64 | 0.62 |
| KCs | 8.111 | 1.90 | 4183 | 67.8 | 8.096[42] | 2.58(30)[22] | 4069.3(15)[42] | 68.394(3)[42] | 8.01 | 1.90 |
| RbCs | 8.380 | 1.21 | 3904 | 49.7 | 8.26[44] | 1.3(1)[51] | 3836.1(5)[43] | 50[47] | 8.28 | 1.24 |

**Table II.** Harmonic vibrational frequencies ($\omega_e$) and anharmonic corrections ($\omega_e\chi_e$) of the XY (X, Y = Li, Na, K, Rb, Cs) heteronuclear alkali dimers. The experimental anharmonic corrections were taken from the same references as the corresponding harmonic frequencies reported in Table I.

|      | This work              |                              | Experiment             |                              | Theory (Ref. 13)       |
|------|------------------------|------------------------------|------------------------|------------------------------|------------------------|
|      | $\omega_e$, cm$^{-1}$ | $\omega_e\chi_e$, cm$^{-1}$ | $\omega_e$, cm$^{-1}$ | $\omega_e\chi_e$, cm$^{-1}$ | $\omega_e$, cm$^{-1}$ |
| LiNa | 257.4                  | 1.8                          | 256.99                 |                              |                        |
| LiK  | 212.4                  | 1.1                          | 211.92(2)              | 1.224                        |                        |
| LiRb | 196.2                  | 1.2                          | 195                    |                              | 185                    |
| LiCs | 182.6                  | 0.7                          | 183                    |                              | 164                    |
| NaK  | 122.4                  | 0.2                          | 124.13                 |                              |                        |
| NaRb | 106.0                  | 0.3                          | 106.965                | 0.3636                       | 107                    |
| NaCs | 97.8                   | 0.3                          | 99                     |                              | 98                     |
| KRb  | 75.3                   | 0.2                          | 75.50                  |                              | 75.5                   |
| KCs  | 67.8                   | 0.2                          | 68.394(3)              | 0.193(1)                     | 66.2                   |
| RbCs | 49.7                   | 0.1                          | 50                     |                              |                        |

**Table III**. Pauling's electronegativity differences ($|X_A-X_B|$), permanent dipole moments ($\mu_e$), harmonic vibrational frequencies ($\omega_e$), and ground vibrational state lifetimes ($\tau$) of the XY (X, Y = Li, Na, K, Rb, Cs) heteronuclear alkali dimers.

| Molecule | $|X_A-X_B|$ | $\mu_e$, D | $\omega_e$, cm$^{-1}$ | Lifetime $\tau$, s | |
|---|---|---|---|---|---|
| | | | | SE solution | Harmonic approx. (Ref. 45) |
| LiNa | 0.05 | 0.54 | 257.4 | $1.3\times10^3$ | - |
| KRb | 0 | 0.65 | 75.3 | $5.6\times10^4$ | $1.3\times10^5$ |
| RbCs | 0.03 | 1.21 | 49.7 | $7.3\times10^4$ | $6.7\times10^4$ |
| KCs | 0.03 | 1.90 | 67.8 | $9.7\times10^3$ | $1.2\times10^4$ |
| NaK | 0.11 | 2.68 | 122.4 | $1.3\times10^3$ | - |
| NaRb | 0.11 | 3.29 | 106.0 | $1.1\times10^3$ | $1.4\times10^3$ |
| LiK | 0.16 | 3.41 | 212.4 | 135 | - |
| LiRb | 0.16 | 4.06 | 196.2 | 102 | 125 |
| NaCs | 0.14 | 4.53 | 97.8 | 542 | 600 |
| LiCs | 0.19 | 5.36 | 182.6 | 53 | 59 |

**Figure captions**

**Figure 1.** Extrapolation of the LiNa dipole moment to the complete basis set limit. The calculations were done at CCSD/cc-pCVXZ (X = D, T, Q, 5) level of theory. The basis set error for quadruple-ζ basis set (ε) is shown.

**Figure 2.** LiNa potential energy curves calculated with different electronic structure methods and the cc-pCVQZ basis set. The experimental values of the dissociation energy and equilibrium distance are indicated with the cross.

**Figure 3.** The potential energy curves (a) and the dipole moment curves (b) of the LiX (X = Na, K, Rb, Cs) molecules calculated with the CCSDT method and quadruple-ζ quality basis sets. The diamonds indicate the experimental values of the dipole moment at the equilibrium internuclear distance $d_{e\ (exp)}$. The squares mark the corresponding theoretical values $d_e$ obtained in this work.

**Figure 4.** Potential energy curves (a) and the dipole moment curves (b) of the NaX (X = Li, K, Rb, Cs) molecules calculated with the CCSDT method and quadruple-ζ quality basis sets. The diamonds indicate the experimental values of the dipole moment at the equilibrium internuclear distance $d_{e\ (exp)}$. The squares mark the corresponding theoretical values $d_e$ obtained in this work.

**Figure 5.** Potential energy curves (a) and the dipole moment curves (b) of the KX (X = Li, Na, Rb, Cs) molecules calculated with the CCSDT method and quadruple-$\zeta$ quality basis sets. The diamonds indicate the experimental values of the dipole moment at the equilibrium internuclear distance $d_{e\ (exp)}$. The squares mark the corresponding theoretical values $d_e$ obtained in this work.

**Figure 6.** Potential energy curves (a) and the dipole moment curves (b) of the RbX (X = Li, Na, K, Cs) molecules calculated with the CCSDT method and quadruple-$\zeta$ quality basis sets. The diamonds indicate the experimental values of the dipole moment at the equilibrium internuclear distance $d_{e\ (exp)}$. The squares mark the corresponding theoretical values $d_e$ obtained in this work.

**Figure 7.** Lifetimes of the vibrational states of the LiX (X = Na, K, Rb, Cs) molecules as functions of the vibration quantum number v.

**Figure 8.** The rates ($\Gamma_i$) of spontaneous emission, and stimulated emission and absorption for the vibrational states of the LiNa molecule. In the insert the transition dipole moments between the initial vibrational levels *i* = 20, 39, 43 and all the final levels *f* < *i* are shown. These transition dipole moments contribute to the Einstein coefficients of spontaneous emission $A_{if}$.

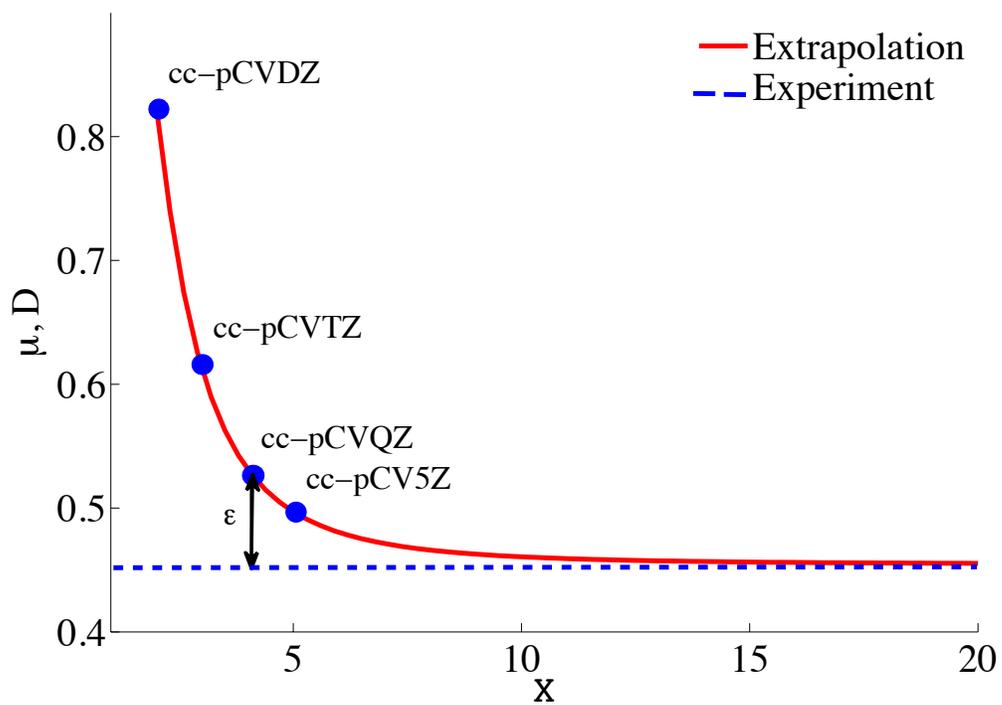

**Figure 1**

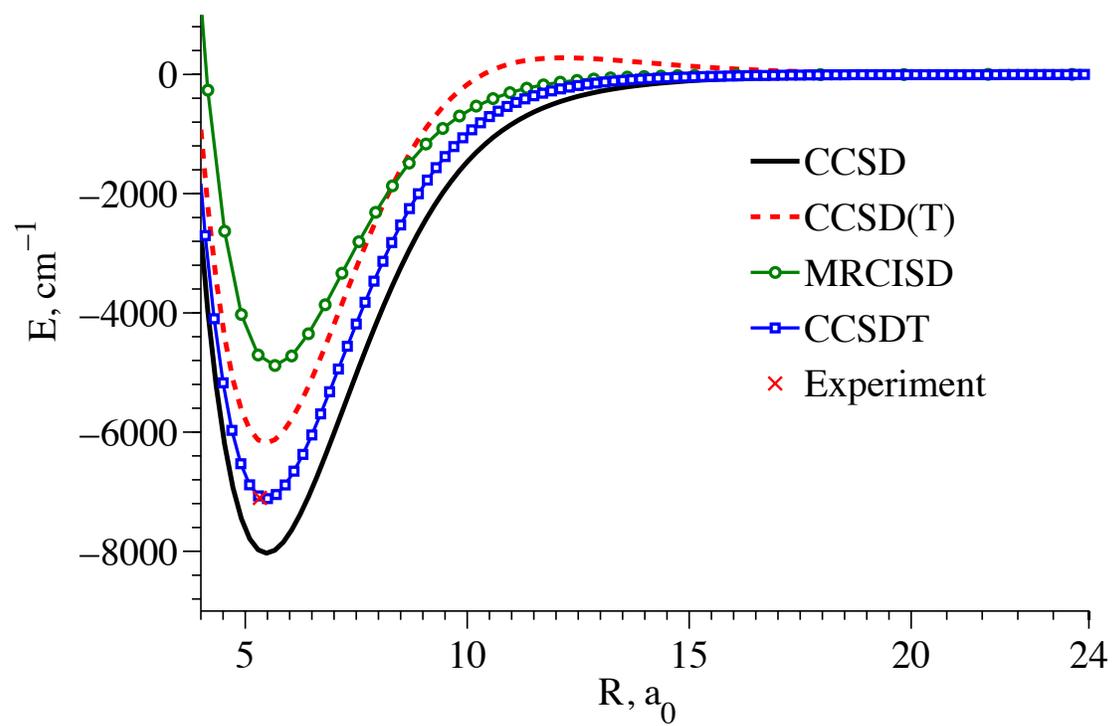

**Figure 2**

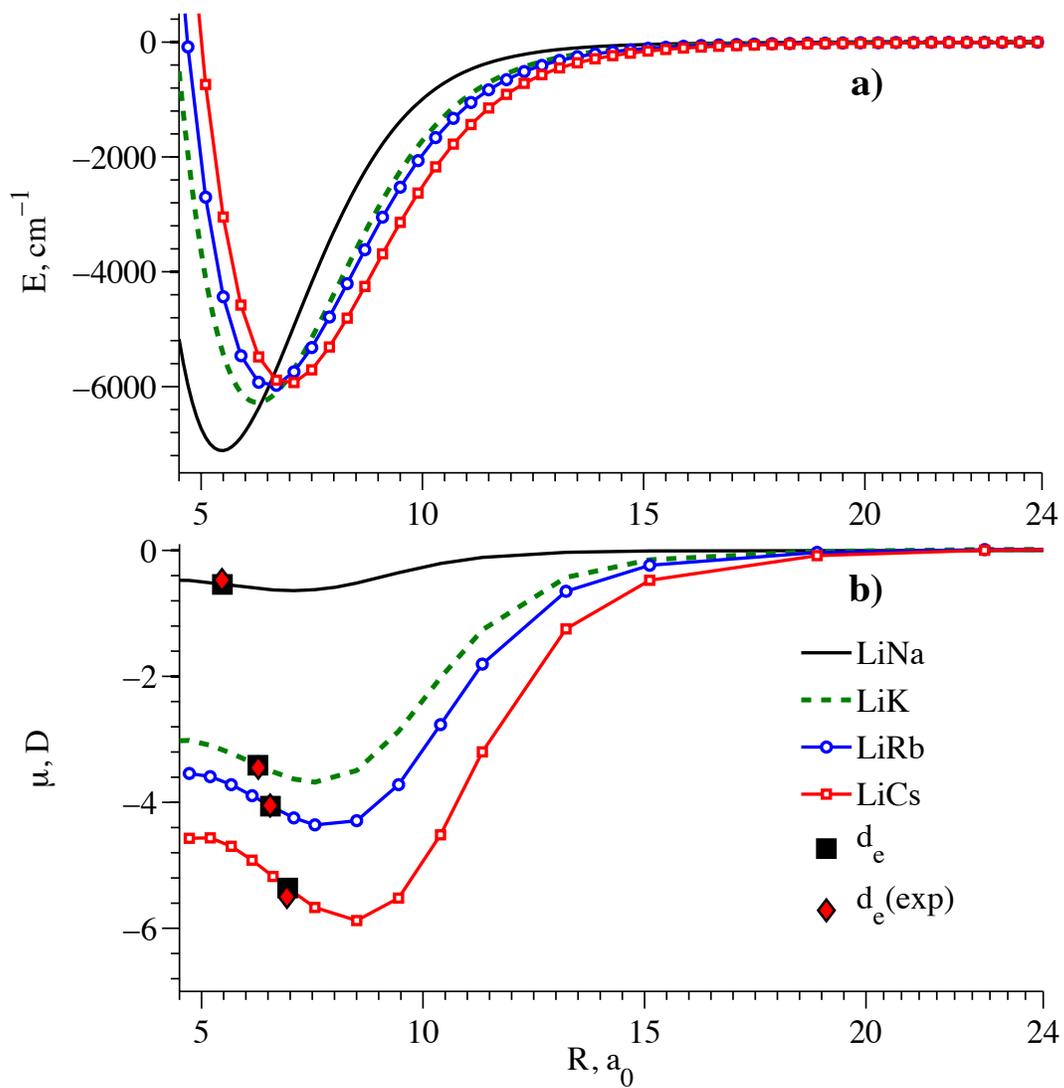

**Figure 3**

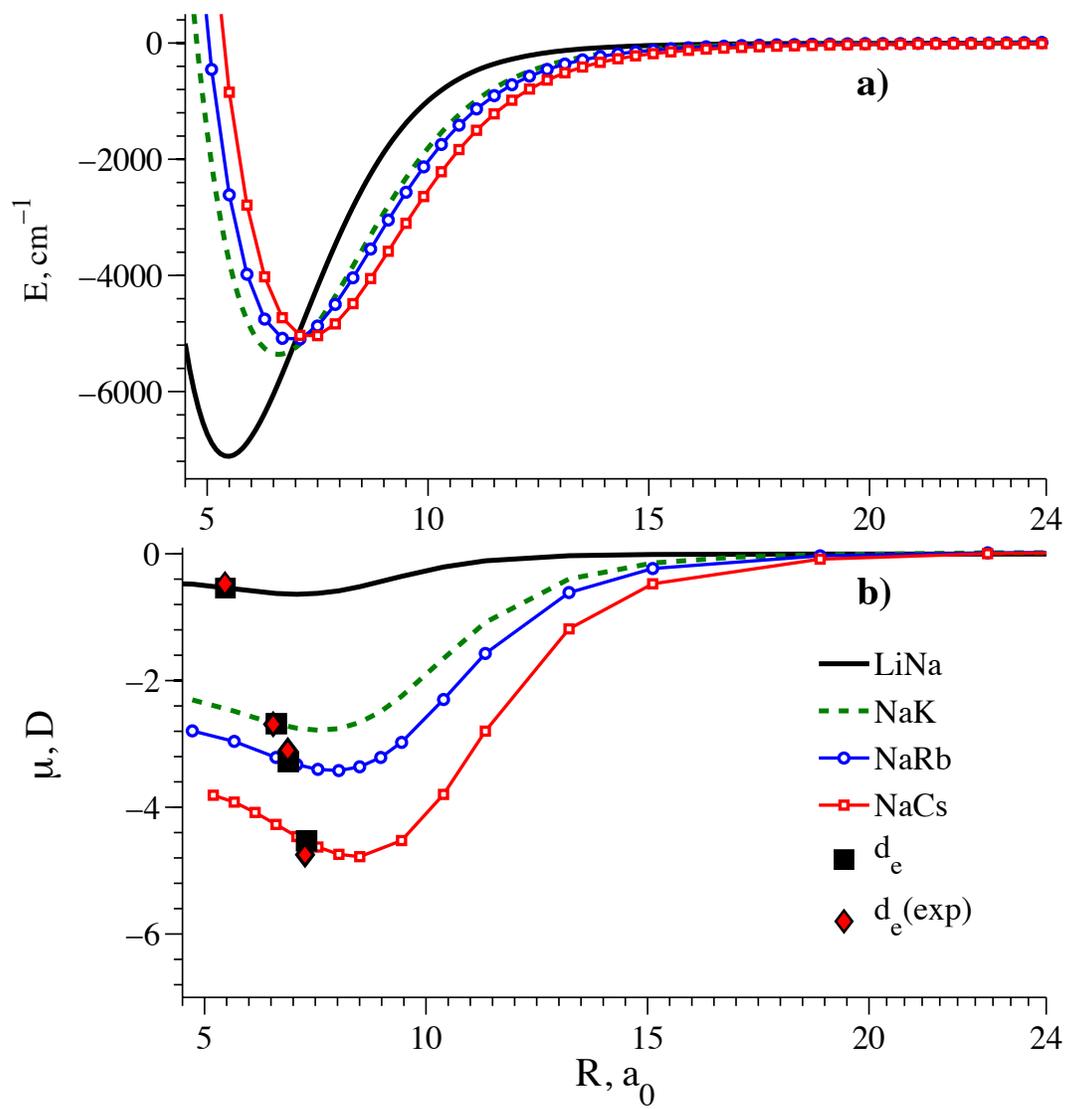

**Figure 4**

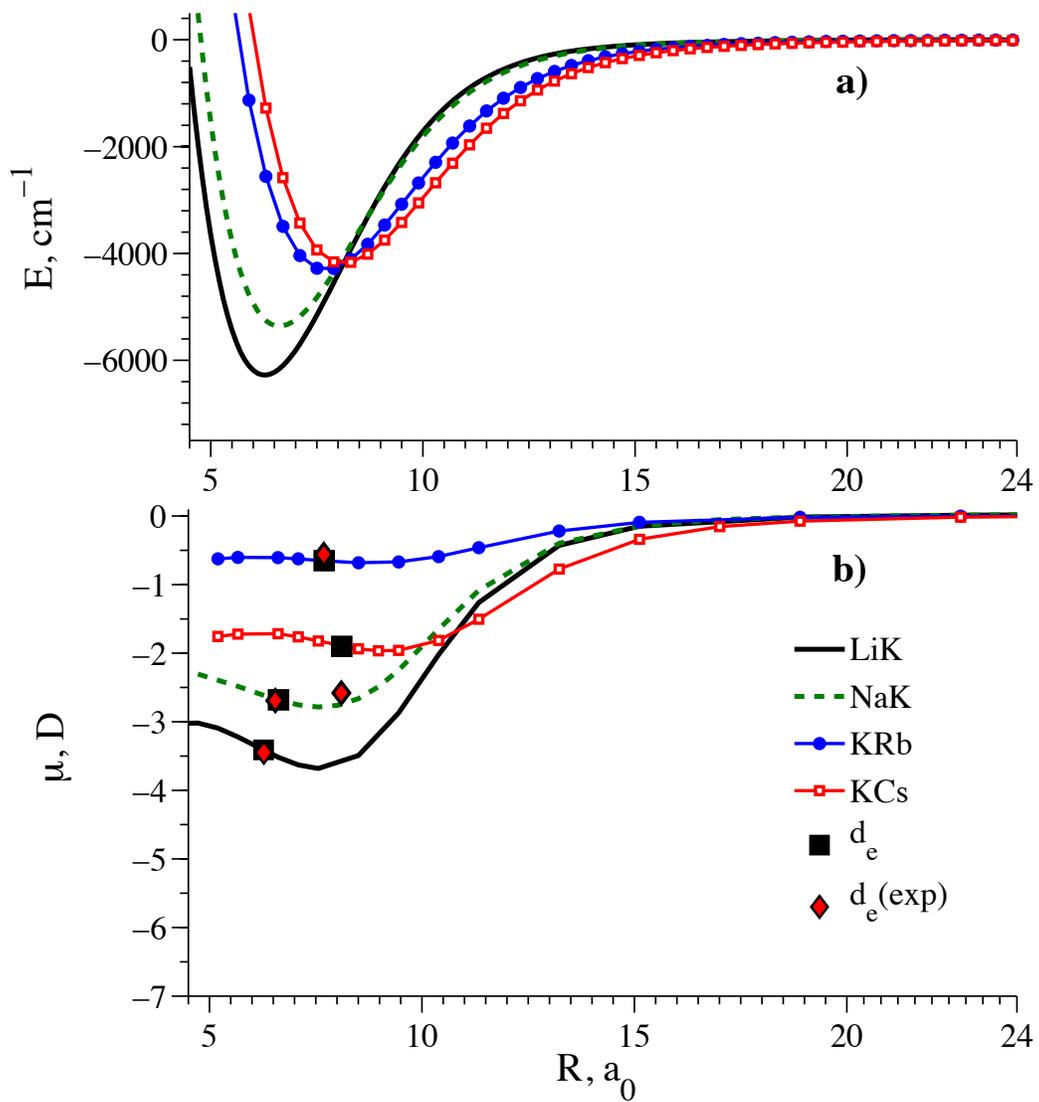

**Figure 5**

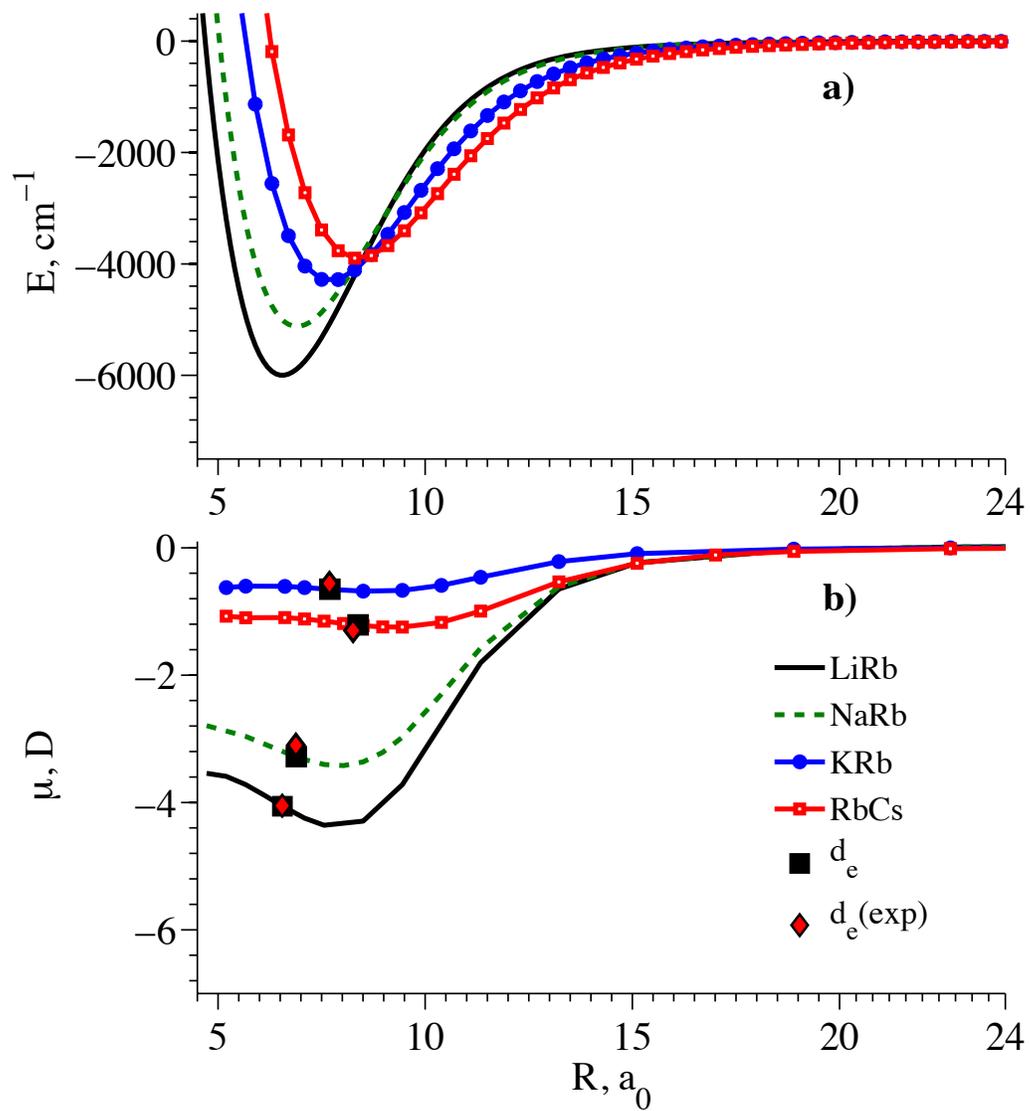

**Figure 6**

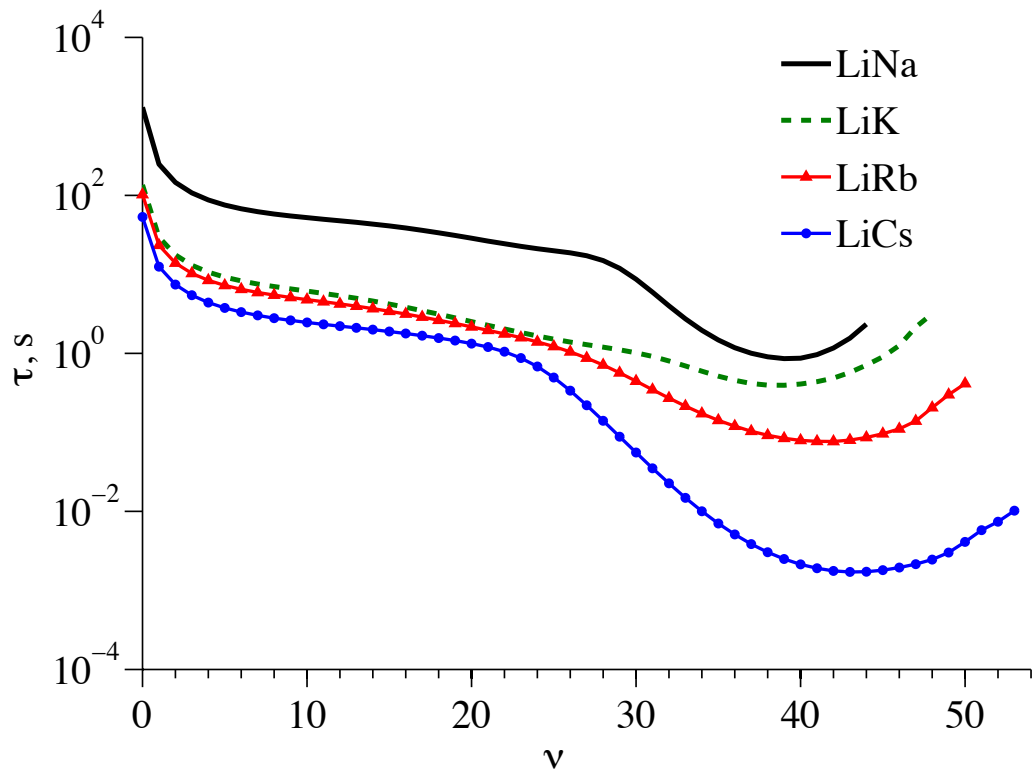

**Figure 7**

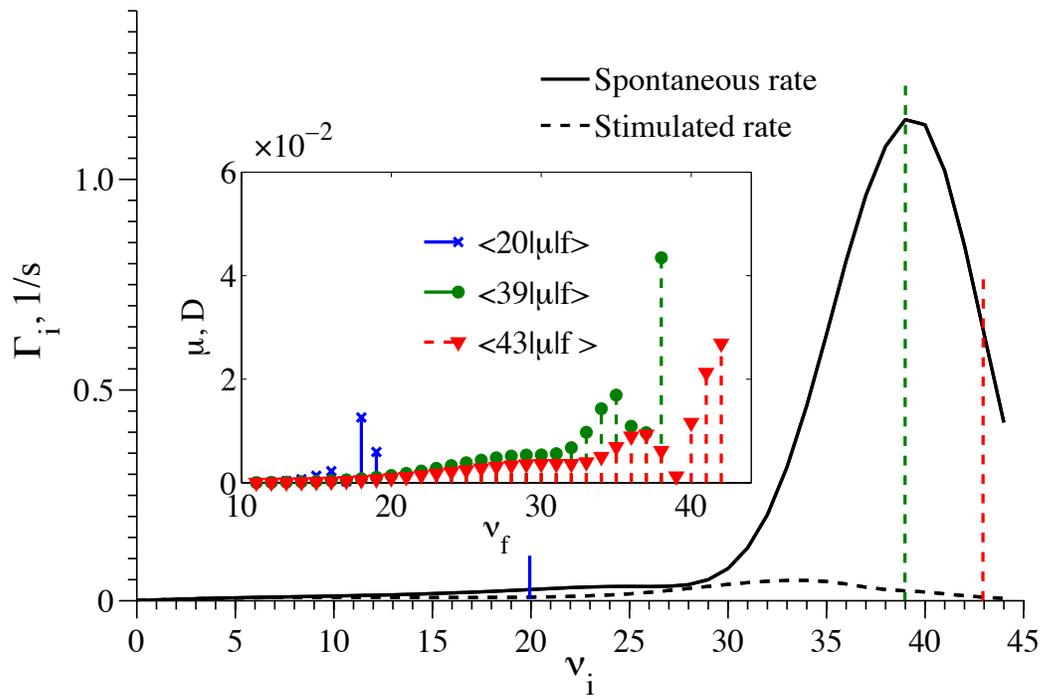

**Figure 8**